\documentclass[sv,article,accept,moreauthors,pdftex]{Definitions/tsp} 
\usepackage[normalem]{ulem} 
\usepackage{amsfonts} 
\usepackage{mathcomp} 
\usepackage{CJKutf8} 
\usepackage{pifont} 
\usepackage{bm} 
\usepackage{bbm} 
\graphicspath{{./Definitions/}} 

\makeatletter
\def\T@n@@nc@d@ngM@cr@M@d{}
\def\LY@n@@nc@d@ngM@cr@M@d{}
\makeatother

\let\orignewcommand\newcommand  
\let\newcommand\providecommand  
\usepackage{verse}
\let\newcommand\orignewcommand  

\newsavebox\foobox

\renewcommand{\dagger}{\mathchar"2279}
\renewcommand{\ddagger}{\mathchar"227A}

\newcommand{\mmathit}[1]{
  \ifthenelse{\equal{#1}{\ln}}{\mathit{ln}}{
    \ifthenelse{\equal{#1}{\max}}{\mathit{max}}{\mathit{#1}}
  }
}
\makeatother
\robustify{\footnote}

\usepackage[braket, qm]{qcircuit}
\usepackage[utf8]{inputenc}
\usepackage[T1]{fontenc}

\DeclareUnicodeCharacter{1E45}{\.{n}}
\DeclareUnicodeCharacter{1E41}{\.{m}}
\DeclareUnicodeCharacter{2003}{\quad}
\DeclareUnicodeCharacter{2009}{\thinspace}
\DeclareUnicodeCharacter{2002}{\enspace{}}
\DeclareUnicodeCharacter{2005}{\thinspace}
\DeclareUnicodeCharacter{0263}{\textipa{G}}
\DeclareUnicodeCharacter{A0}{~}
\DeclareUnicodeCharacter{2460}{\textcircled{\scriptsize{1}}}
\DeclareUnicodeCharacter{2461}{\textcircled{\scriptsize{2}}}
\DeclareUnicodeCharacter{2462}{\textcircled{\scriptsize{3}}}
\DeclareUnicodeCharacter{2463}{\textcircled{\scriptsize{4}}}
\DeclareUnicodeCharacter{2464}{\textcircled{\scriptsize{5}}}
\DeclareUnicodeCharacter{2465}{\textcircled{\scriptsize{6}}}
\DeclareUnicodeCharacter{2466}{\textcircled{\scriptsize{7}}}
\DeclareUnicodeCharacter{2467}{\textcircled{\scriptsize{8}}}
\DeclareUnicodeCharacter{2468}{\textcircled{\scriptsize{9}}}
\DeclareUnicodeCharacter{2070}{\textsuperscript{0}}
\DeclareUnicodeCharacter{2074}{\textsuperscript{4}}
\DeclareUnicodeCharacter{2075}{\textsuperscript{5}}
\DeclareUnicodeCharacter{2076}{\textsuperscript{6}}
\DeclareUnicodeCharacter{2077}{\textsuperscript{7}}
\DeclareUnicodeCharacter{2078}{\textsuperscript{8}}
\DeclareUnicodeCharacter{2079}{\textsuperscript{9}}
\DeclareUnicodeCharacter{02C2}{<}
\DeclareUnicodeCharacter{2033}{\relax\ifmmode '' \else $''$\fi}
\DeclareUnicodeCharacter{2034}{\relax\ifmmode ''' \else $'''$\fi}
\DeclareUnicodeCharacter{2026}{\relax\ifmmode … \else $\ldots$\fi}
\DeclareUnicodeCharacter{0229}{\c{e}}
\DeclareUnicodeCharacter{016F}{\r{u}}
\DeclareUnicodeCharacter{127}{\relax\ifmmode\rm\hbar\else $\rm\hbar$\fi}
\DeclareUnicodeCharacter{3AC}{\relax\ifmmode\acute{\alpha}\else $\acute{\alpha}$\fi}
\DeclareUnicodeCharacter{3AD}{\relax\ifmmode\acute{\varepsilon}\else $\acute{\varepsilon}$\fi}
\DeclareUnicodeCharacter{3AE}{\relax\ifmmode\acute{\eta}\else $\acute{\eta}$\fi}
\DeclareUnicodeCharacter{3AF}{\relax\ifmmode\acute{\iota}\else $\acute{\iota}$\fi}
\DeclareUnicodeCharacter{3CC}{\relax\ifmmode\acute{o}\else $\acute{o}$\fi}
\DeclareUnicodeCharacter{3CD}{\relax\ifmmode\acute{\upsilon}\else $\acute{\upsilon}$\fi}
\DeclareUnicodeCharacter{3CE}{\relax\ifmmode\acute{\omega}\else $\acute{\omega}$\fi}
\DeclareUnicodeCharacter{391}{A}
\DeclareUnicodeCharacter{392}{B}
\DeclareUnicodeCharacter{395}{E}
\DeclareUnicodeCharacter{396}{Z}
\DeclareUnicodeCharacter{397}{H}
\DeclareUnicodeCharacter{399}{I}
\DeclareUnicodeCharacter{39A}{K}
\DeclareUnicodeCharacter{39C}{M}
\DeclareUnicodeCharacter{39D}{N}
\DeclareUnicodeCharacter{39F}{O}
\DeclareUnicodeCharacter{3A1}{P}
\DeclareUnicodeCharacter{3A4}{T}
\DeclareUnicodeCharacter{3A7}{X}

\DeclareUnicodeCharacter{27E6}{\relax\ifmmode \llbracket \else $\llbracket$\fi}
\DeclareUnicodeCharacter{27E7}{\relax\ifmmode \rrbracket \else $\rrbracket$\fi}

\DeclareUnicodeCharacter{1D434}{\relax\ifmmode A \else $A$\fi}
\DeclareUnicodeCharacter{1D435}{\relax\ifmmode B \else $B$\fi}
\DeclareUnicodeCharacter{1D436}{\relax\ifmmode C \else $C$\fi}
\DeclareUnicodeCharacter{1D437}{\relax\ifmmode D \else $D$\fi}
\DeclareUnicodeCharacter{1D438}{\relax\ifmmode E \else $E$\fi}
\DeclareUnicodeCharacter{1D439}{\relax\ifmmode F \else $F$\fi}
\DeclareUnicodeCharacter{1D43A}{\relax\ifmmode G \else $G$\fi}
\DeclareUnicodeCharacter{1D43B}{\relax\ifmmode H \else $H$\fi}
\DeclareUnicodeCharacter{1D43C}{\relax\ifmmode I \else $I$\fi}
\DeclareUnicodeCharacter{1D43D}{\relax\ifmmode J \else $J$\fi}
\DeclareUnicodeCharacter{1D43E}{\relax\ifmmode K \else $K$\fi}
\DeclareUnicodeCharacter{1D43F}{\relax\ifmmode L \else $L$\fi}
\DeclareUnicodeCharacter{1D440}{\relax\ifmmode M \else $M$\fi}
\DeclareUnicodeCharacter{1D441}{\relax\ifmmode N \else $N$\fi}
\DeclareUnicodeCharacter{1D442}{\relax\ifmmode O \else $O$\fi}
\DeclareUnicodeCharacter{1D443}{\relax\ifmmode P \else $P$\fi}
\DeclareUnicodeCharacter{1D444}{\relax\ifmmode Q \else $Q$\fi}
\DeclareUnicodeCharacter{1D445}{\relax\ifmmode R \else $R$\fi}
\DeclareUnicodeCharacter{1D446}{\relax\ifmmode S \else $S$\fi}
\DeclareUnicodeCharacter{1D447}{\relax\ifmmode T \else $T$\fi}
\DeclareUnicodeCharacter{1D448}{\relax\ifmmode U \else $U$\fi}
\DeclareUnicodeCharacter{1D449}{\relax\ifmmode V \else $V$\fi}
\DeclareUnicodeCharacter{1D44A}{\relax\ifmmode W \else $W$\fi}
\DeclareUnicodeCharacter{1D44B}{\relax\ifmmode X \else $X$\fi}
\DeclareUnicodeCharacter{1D44C}{\relax\ifmmode Y \else $Y$\fi}
\DeclareUnicodeCharacter{1D44D}{\relax\ifmmode Z \else $Z$\fi}
\DeclareUnicodeCharacter{1D44E}{\relax\ifmmode a \else $a$\fi}
\DeclareUnicodeCharacter{1D44F}{\relax\ifmmode b \else $b$\fi}
\DeclareUnicodeCharacter{1D450}{\relax\ifmmode c \else $c$\fi}
\DeclareUnicodeCharacter{1D451}{\relax\ifmmode d \else $d$\fi}
\DeclareUnicodeCharacter{1D452}{\relax\ifmmode e \else $e$\fi}
\DeclareUnicodeCharacter{1D453}{\relax\ifmmode f \else $f$\fi}
\DeclareUnicodeCharacter{1D454}{\relax\ifmmode g \else $g$\fi}
\DeclareUnicodeCharacter{1D456}{\relax\ifmmode i \else $i$\fi}
\DeclareUnicodeCharacter{1D457}{\relax\ifmmode j \else $j$\fi}
\DeclareUnicodeCharacter{1D458}{\relax\ifmmode k \else $k$\fi}
\DeclareUnicodeCharacter{1D459}{\relax\ifmmode l \else $l$\fi}
\DeclareUnicodeCharacter{1D45A}{\relax\ifmmode m \else $m$\fi}
\DeclareUnicodeCharacter{1D45B}{\relax\ifmmode n \else $n$\fi}
\DeclareUnicodeCharacter{1D45C}{\relax\ifmmode o \else $o$\fi}
\DeclareUnicodeCharacter{1D45D}{\relax\ifmmode p \else $p$\fi}
\DeclareUnicodeCharacter{1D45E}{\relax\ifmmode q \else $q$\fi}
\DeclareUnicodeCharacter{1D45F}{\relax\ifmmode r \else $r$\fi}
\DeclareUnicodeCharacter{1D460}{\relax\ifmmode s \else $s$\fi}
\DeclareUnicodeCharacter{1D461}{\relax\ifmmode t \else $t$\fi}
\DeclareUnicodeCharacter{1D462}{\relax\ifmmode u \else $u$\fi}
\DeclareUnicodeCharacter{1D463}{\relax\ifmmode v \else $v$\fi}
\DeclareUnicodeCharacter{1D464}{\relax\ifmmode w \else $w$\fi}
\DeclareUnicodeCharacter{1D465}{\relax\ifmmode x \else $x$\fi}
\DeclareUnicodeCharacter{1D466}{\relax\ifmmode y \else $y$\fi}
\DeclareUnicodeCharacter{1D467}{\relax\ifmmode z \else $z$\fi}

\DeclareUnicodeCharacter{1E67}{\.{\v s}}
\DeclareUnicodeCharacter{1E11}{\relax\ifmmode \c{d} \else $\c{d}$\fi}
\DeclareUnicodeCharacter{1ECB}{\relax\ifmmode \d{i} \else $\d{i}$\fi}
\DeclareUnicodeCharacter{1D8D}{\relax\ifmmode \textlhookx \else $\textlhookx$\fi}
\DeclareUnicodeCharacter{104}{\relax\ifmmode \k{A} \else $\k{A}$\fi}
\DeclareUnicodeCharacter{211E}{\relax\ifmmode \textrecipe \else $\textrecipe$\fi}
\DeclareUnicodeCharacter{29D}{\relax\ifmmode \textctj \else $\textctj$\fi}

\DeclareUnicodeCharacter{1E2E}{\'{\"I}}
\DeclareUnicodeCharacter{23F}{\textrts}

\DeclareUnicodeCharacter{2C73}{\varw}

\DeclareUnicodeCharacter{2127}{\mho}

\DeclareUnicodeCharacter{28C}{\textturnv}
\DeclareUnicodeCharacter{252}{\textturnscripta}
\DeclareUnicodeCharacter{259}{\schwa}
\DeclareUnicodeCharacter{25B}{\m{e}}
\DeclareUnicodeCharacter{266}{\m{h}}
\DeclareUnicodeCharacter{127}{\B{h}}
\DeclareUnicodeCharacter{27E}{\textfishhookr}
\DeclareUnicodeCharacter{281}{\textinvscr}

\continuouspages{yes}
\firstpage{1} 
\pubvolume{1}
\issuenum{1}
\articlenumber{12345}
\pubyear{2025}
\copyrightyear{2025}
\dateonlinefirst{}
\datepublished{}

\Title{From Qubits to Rhythm: Exploring Quantum Random Walks in Rhythmspaces}

\Author{María Aguado-Yáñez\textsuperscript{1,4,*}\orcidA, Karl Jansen\textsuperscript{2}, Daniel Gómez-Marín\textsuperscript{3} and Sergi Jordà\textsuperscript{4}\orcidD}

\AuthorNames{María Aguado-Yáñez, Karl Jansen, Daniel Gómez-Marín and Sergi Jordà}

\address{%
\textsuperscript{1}Interdisciplinary Centre for Computer Music Research, University of Plymouth,  Plymouth, PL48AA, United Kingdom

\textsuperscript{2}Centre for Quantum Technologies and Applications, Deutsches Elektronen-Synchrotron DESY, Zeuthen, 15738, Germany

\textsuperscript{3}Universidad Icesi, Cali, Colombia

\textsuperscript{4}Music Technology Group, Pompeu Fabra University, Barcelona, 08018, Spain
}

\corres{Corresponding Author: María Aguado-Yáñez. Email: maria.aguadoyanez@plymouth.ac.uk}

\firstnote{} 
\secondnote{}

\abstract{A quantum computing algorithm for rhythm generation is presented, which aims to expand and explore quantum computing applications in the arts, particularly in music. The algorithm maps quantum random walk trajectories onto a rhythmspace -- a 2D interface that interpolates rhythmic patterns. The methodology consists of three stages. The first stage involves designing quantum computing algorithms and establishing a mapping between the qubit space and the rhythmspace. To minimize circuit depth, a decomposition of a 2D quantum random walk into two 1D quantum random walks is applied. The second stage focuses on biasing the directionality of quantum random walks by introducing classical potential fields, adjusting the probability distribution of the wave function based on the position gradient within these fields. Four potential fields are implemented: a null potential, a linear field, a Gaussian potential, and a Gaussian potential under inertial dynamics. The third stage addresses the sonification of these paths by generating MIDI drum pattern messages and transmitting them to a Digital Audio Workstation (DAW). This work builds upon existing literature that applies quantum computing to simpler qubit spaces with a few positions, extending the formalism to a 2D x-y plane. It serves as a proof of concept for scalable quantum computing-based generative random walk algorithms in music and audio applications. Furthermore, the approach is applicable to generic multidimensional sound spaces, as the algorithms are not strictly constrained to rhythm generation and can be adapted to different musical structures.}


\begin{document}

\section{Introduction} \label{sect:intro}
As quantum computing continues to expand into disciplines such as cryptography and machine learning \cite{quantumcomputing2025}, this work explores and expands quantum computing algorithms for music and audio applications \cite{miranda2022, miranda2024}. The goal is to develop proof-of-concept, scalable, and robust quantum random walk algorithms for rhythmic pattern generation, providing a foundation for more complex techniques to be built upon as the fault-tolerant era of quantum computers is reached. The tool used for sonification is a rhythmspace, a 2D interface-based rhythm generative tool.

Since this paper bridges quantum random walks, and rhythmspaces, both concepts are briefly introduced here. A quantum random walk \cite{venegas2012} is a trajectory model in quantum physics which incorporates the principles of quantum mechanics into the path-generation process (see \cite{feynman1965} for an introductory lecture on Quantum Mechanics). It is the quantum counterpart of a classical random walk: instead of moving randomly between positions, a quantum walker evolves through superposition and interference. This allows the particle to exist in multiple states simultaneously, with its movement governed by probability amplitudes, resulting in a very different spreading behavior compared to its classical analogue. When the states of a quantum random walk are encoded using qubits, the process can be implemented as a quantum circuit using the gate-based formalism of quantum computing. In this work, a two-dimensional discrete quantum random walk is used, where the walker moves along two axes, matching the dimensionality of the rhythmspace.

A rhythmspace \cite{turquois2016, gomezmarin2018, gomezmarin2020} is a tool that allows users to generate rhythmic patterns by clicking and dragging on a two-dimensional visual interface. In music production, DJs, producers, and musicians often face the challenge of sifting through large sound libraries to find suitable rhythms. Rhythmspaces offer a novel, intuitive, and visually organized solution to this problem, allowing users to explore and select rhythms more efficiently. By clicking on different points in the interface, users can navigate this space of rhythms and immediately hear the corresponding pattern. A demonstration video of the tool can be found in \cite{gomezmarindemo2018}. In this paper, quantum random walk paths are mapped onto a rhythmspace, causing the rhythmic patterns to evolve dynamically in response to the walk’s probabilistic nature.

There are only a few references that have used quantum computing random walk algorithms for music generation. In particular, two chapters within \cite{miranda2022} explore the concept of quantum random walks. In the first one \cite{mirandabasak2022}, the notion of a quantum random walk is introduced deriving from classical models such as classical random walks and Markov chains. A system of 3 qubits is then mapped onto a cube to design a simple quantum walk through its vertices. In the second one \cite{allen2022}, discrete quantum random walks are described in more detail and used to generate basic musical sequences. 

This work aims to extend the content of those chapters. Firstly, following the approach in the first chapter \cite{mirandabasak2022} and drawing inspiration from the cube-vertex encoding, a more complex geometry is proposed: a discrete x-y grid composed of a significantly larger number of positions. The decision to scale up to a 2D grid is motivated by the structure of the rhythmspaces interface, which follows this layout. Secondly, building upon the formalism in the second chapter \cite{allen2022}, an extended mathematical framework is developed for larger systems of qubits arranged as quantum random walks, specifically adapted to the proposed x-y grid. This provides a proof of concept and a scalable protocol for using qubits as quantum random walks in higher-dimensional musical spaces.

Designing quantum computing systems for artistic and musical applications is important to expand the creative possibilities within artistic processes and to broaden the reach of science in society. The properties inherent to quantum mechanics are unique and cannot be reproduced by classical physics, and new types of tools and algorithms can be developed to produce art in innovative ways and to exemplify quantum physics to audiences of all levels. Sonifying abstract mathematical spaces and physical systems can help the research community perceive mathematics and physics from alternative perspectives and gain deeper insights. Music, being a highly popular art form and an essential part of all cultures, can support this understanding. Exploring physics through intuitive, perceptual approaches—similar to creating visual diagrams or graphs—is a powerful tool for scientists. Finally, making quantum formalism more accessible to the general public through music can lead to innovative teaching methods, transforming dense lectures on quantum mechanics into interactive auditory experiences. Papers \cite{yamada2023, crippa2024} exemplify these ideas well. In the former \cite{yamada2023}, examples of sonification of quantum systems are presented. In the latter \cite{crippa2024}, classical paintings are reinterpreted using quantum algorithms as an example of quantum computing applied to visual art.

\begin{figure}[H]
    \centering
    \includegraphics[scale=0.8]{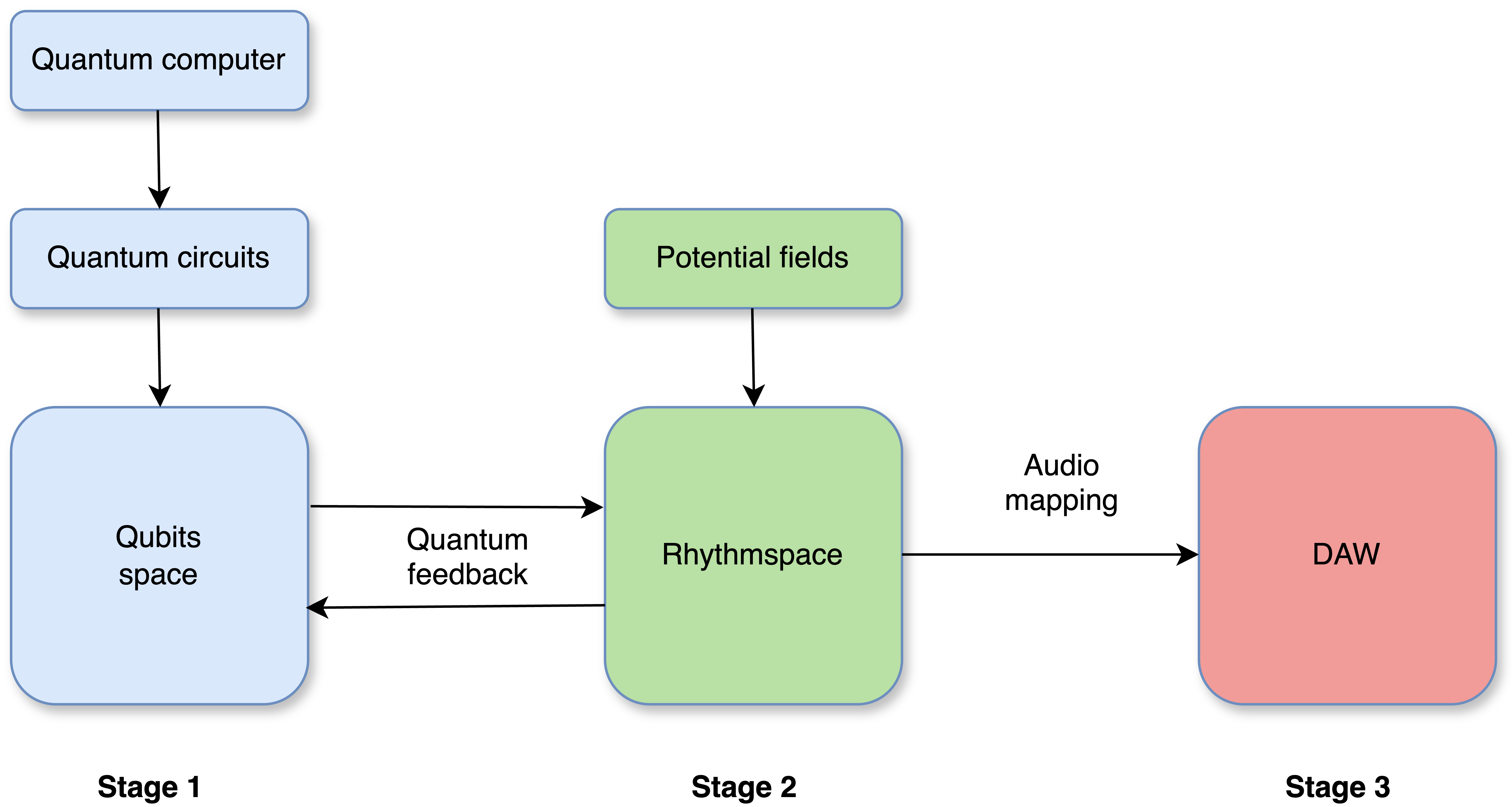}
    \caption{Pipeline of the mappings and information exchange between the components of the code.}
    \label{fig:stages}
\end{figure}

The process carried out for this work is divided into three main stages, which define the structure of this paper. See Fig. \ref{fig:stages} for a visual overview of the workflow. Section 2 covers the first stage, where quantum random walk algorithms are designed and mapped onto the rhythmspace. Section 3 presents the second stage, explaining how classical potential fields are introduced into the rhythmspace to bias the generation of quantum random walk paths, guiding them to follow desired trajectories. This interaction between the rhythmspace potential fields and the quantum walk generation in the qubit space is referred to as quantum feedback. Section 4 addresses the third stage, which involves sonification: the generated paths are transformed into MIDI messages and sent to a DAW (Digital Audio Workstation) to be converted to audio. The entire codebase for this work is written in Python, and Qiskit is used to construct and run the quantum circuits. Finally, Section 5 covers the discussion and final conclusions of this work. 

\section{First stage: mapping the qubit space to the rhythmspace} \label{sect:stage1}

This first stage addresses how quantum random walk algorithms are designed and mapped onto the rhythmspace. To begin with, the qubit Hilbert spaces used to encode the walks are defined. Then, the circuit that implements the walk is constructed to run on the previously established Hilbert spaces. Finally, the mapping from the Hilbert spaces onto the rhythmspace is specified. These steps are explored in the following three subsections within this stage.

\subsection{Defining the qubit Hilbert spaces}
The definition of the qubit Hilbert spaces should be guided by the mapping between the qubit space and the rhythmspace, which must be designed to support the representation of quantum random walks within the rhythmspaces' 2D interface. This mapping is derived from the formalism of a 1D quantum random walk. A quantum random walk is composed of two Hilbert spaces: the position space, $\mathcal{H}_p$, and the coin space, $\mathcal{H}_c$, as shown in Eq. (\ref{eq:1d-spaces}):

\begin{align}
    \mathcal{H}_{p}^{1D} &= \text{span}\{ \ket{x}, x \in \mathbb{Z}_N  \}, \notag\\
    \mathcal{H}_{c}^{1D} &= \text{span}\{ \ket{\rightarrow}, \ket{\leftarrow} \}, \notag\\
    \label{eq:1d-spaces}
\end{align}

where $\ket{x}$ are the positions the walker can take, and $\ket{\rightarrow}, \ket{\leftarrow}$ are the heads and tails possible outputs of the coin. To advance one step in a 1D quantum random walk, the coin is tossed, and the walker takes a step forward or backward depending on its output. The total space is composed as defined in Eq. (\ref{eq:total-hilbert-1d}):

\begin{equation}
     \mathcal{H}^{1D} = \mathcal{H}_p^{1D} \otimes \mathcal{H}_c^{1D}.
     \label{eq:total-hilbert-1d}
\end{equation}

In a 1D quantum random walk, the position space $\mathcal{H}_p^{1D}$ can be encoded with $\log_2(N)$ qubits, where $N$ is the dimension of the space.

Performing the operation of tossing the coin means applying an operator, such as the one shown in Eq. (\ref{eq:coin-op-1d}), to a given initial state, introducing a superposition of the heads and tails states:

\begin{equation}
     C^{1D} = \frac{1}{\sqrt{2}}
     \left(
     \ket{\rightarrow} \bra{\rightarrow} + \ket{\rightarrow} \bra{\leftarrow} + \ket{\leftarrow} \bra{\rightarrow} - \ket{\leftarrow}  \bra{\leftarrow} \right).
     \label{eq:coin-op-1d}
\end{equation}

The next operator applied, given in Eq. (\ref{eq:position-op-1d}), is the one that performs a shift in position depending on the superposition of the coin:

\begin{equation}
     S^{1D} = \sum_{x} \left( |x+1\rangle \langle x| \otimes  \ket{\rightarrow} \bra{\rightarrow} + |x-1\rangle \langle x| \otimes \ket{\leftarrow}  \bra{\leftarrow} \right).
    \label{eq:position-op-1d}
\end{equation}

Finally, Eq. \eqref{eq:evolution-operator-1d} shows that the overall evolution is described by the concatenation of the coin-tossing operator and the shift operator:

\begin{equation}
     U^{1D} = S^{1D} \cdot (C^{1D} \otimes I ).
    \label{eq:evolution-operator-1d}
\end{equation}

To generate the path of a 1D quantum random walk, this operator must be applied as many times as steps are taken before measuring the wave function.

The 2D quantum random walk formalism is obtained from the 1D case. A 2D walk is needed because the dimensionality of the rhythmspace is 2D as well. The Hilbert spaces involved in a 2D quantum random walk are defined in Eq. (\ref{eq:2d-spaces}):

\begin{align}
    \mathcal{H}_{p}^{2D} &= \text{span}\{ \ket{x,y}, x,y \in \mathbb{Z}_N  \}, \notag\\
    \mathcal{H}_{c}^{2D} &= \text{span}\{\ket{\nearrow}, \ket{\nwarrow}, \ket{\searrow},
    \ket{\swarrow}\}, \notag\\
     \mathcal{H}^{2D} &= \mathcal{H}_p^{2D} \otimes \mathcal{H}_c^{2D},
    \label{eq:2d-spaces}
\end{align}

where the states $\ket{x,y}$ encode the positions of a 2D grid, and the states $\ket{\nearrow}, \ket{\nwarrow}, \ket{\searrow}, \ket{\swarrow}$ encode the four possible outputs of a coin in a plane. In this case, $\mathcal{H}_p^{2D}$ can be encoded using $\log_2(N^2) = 2\log_2(N)$ qubits. The evolution operator for a 2D quantum random walk, shown in Eq. (\ref{eq:evolution-operator-2d}), is given by:

\begin{equation}
     U^{2D} = S^{2D} \cdot (C^{2D} \otimes I ).
    \label{eq:evolution-operator-2d}
\end{equation}

A total of 8 qubits are used, dedicating 6 to the position space and 2 to the coin space. The mapping chosen between the qubits and the positions of the rhythmspace is given in Eq. \eqref{eq:mapping-positions}:

\begin{align}
    \ket{x,y}_p &= \ket{q_1 q_2 q_3, q_4 q_5 q_6}, \notag\\
    x &= 2^2 q_1 + 2^1 q_2 + 2^0 q_3, \notag\\
    y &= 2^2 q_4 + 2^1 q_5 + 2^0 q_6,
    \label{eq:mapping-positions}
\end{align}

and the mapping for the coins is the one in Eq. \eqref{eq:mapping-coins}:

\begin{equation}
    \ket{x,y}_c = \ket{q_7, q_8}.
    \label{eq:mapping-coins}
\end{equation}

The position qubits are separated into two groups of three, each encoding one axis of the rhythmspace. This arrangement keeps the two axes separated, allowing the total Hilbert space to be expressed as a composition of two one-dimensional quantum random walks, as shown in Eq. \eqref{eq:decomposition}:

\begin{equation}
    \mathcal{H}^{2D} \cong \mathcal{H}_x^{1D} \otimes \mathcal{H}_y^{1D},
    \label{eq:decomposition}
\end{equation}

where the isomorphism is given by a swap of qubits:

\begin{equation}
     U^{2D} = S^{2D} \cdot (C^{2D} \otimes I ) \cong U^{1D} \otimes U^{1D} = S_x^{1D}(C_x^{1D} \otimes I_x) \otimes S_y^{1D}(C_y^{1D}\otimes I_y).
    \label{eq:isomorphism}
\end{equation}

In other words, Eq. (\ref{eq:isomorphism}) shows that it is indeed possible to decompose a 2D quantum random walk into two 1D walks simply by rearranging the qubits. This decomposition is relevant because it enables a circuit structure with low depth. Carefully selecting the operators and Hilbert spaces involved—with the aim of reducing the number of qubits later on—is a key aspect in the design of quantum computing algorithms.

\subsection{Crafting the circuit}
Once the spaces and operators are defined, the next step is to construct the circuit. At this stage, the decomposition into one-dimensional axes becomes crucial. The initial approach consisted of simply extending the matrices for a one-dimensional walk from \cite{allen2022} to the 2D case. However, when these matrices were inserted into the \texttt{transpile()} function in Qiskit to decompose them into a set of universal gates, the resulting circuit contained thousands of gates. Applying an open-source algorithm from \cite{dmytrofedoriaka2020} reduced the number of gates from thousands to hundreds compared to the previous approach, but the reduction was still insufficient. Decomposing a large operator into a set of simple quantum gates can be highly demanding in terms of quantum computational resources. A more efficient approach is to account for an algebraic representation of the matrices that minimizes the number of gates required. If the process previously applied to the 1D case is followed without considering this aspect, the resulting matrix could again require thousands of gates, whereas a practical implementation typically keeps this number under 100. If the 2D quantum walk is decomposed into two 1D quantum walks, a simple circuit implementation with a small number of gates from \cite{venegas2023} can be used. The matrix representations and decomposition methods are explained in detail there, and the proof that a 2D walk can be composed of two 1D walks was provided in Eq. (\ref{eq:isomorphism}). Fig. \ref{fig:my-circuit} shows the resulting circuit:

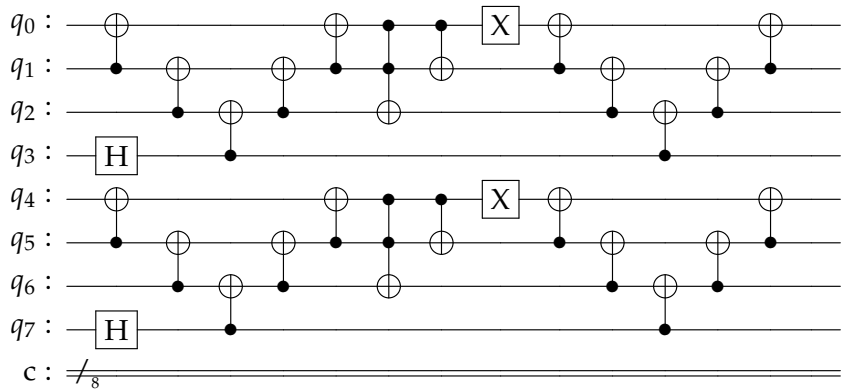
\begin{figure}[h!]
    \centering
    \scalebox{1.0}{
    \Qcircuit @C=1.0em @R=0.2em @!R {
        \nghost{{q}_{0} :  } & \lstick{{q}_{0} :  } & \targ & \qw & \qw & \qw & \targ & \ctrl{1} & \ctrl{1} & \gate{\mathrm{X}} & \targ & \qw & \qw & \qw & \targ & \qw & \qw\\
        \nghost{{q}_{1} :  } & \lstick{{q}_{1} :  } & \ctrl{-1} & \targ & \qw & \targ & \ctrl{-1} & \ctrl{1} & \targ & \qw & \ctrl{-1} & \targ & \qw & \targ & \ctrl{-1} & \qw & \qw\\
        \nghost{{q}_{2} :  } & \lstick{{q}_{2} :  } & \qw & \ctrl{-1} & \targ & \ctrl{-1} & \qw & \targ & \qw & \qw & \qw & \ctrl{-1} & \targ & \ctrl{-1} & \qw & \qw & \qw\\
        \nghost{{q}_{3} :  } & \lstick{{q}_{3} :  } & \gate{\mathrm{H}} & \qw & \ctrl{-1} & \qw & \qw & \qw & \qw & \qw & \qw & \qw & \ctrl{-1} & \qw & \qw & \qw & \qw\\
        \nghost{{q}_{4} :  } & \lstick{{q}_{4} :  } & \targ & \qw & \qw & \qw & \targ & \ctrl{1} & \ctrl{1} & \gate{\mathrm{X}} & \targ & \qw & \qw & \qw & \targ & \qw & \qw\\
        \nghost{{q}_{5} :  } & \lstick{{q}_{5} :  } & \ctrl{-1} & \targ & \qw & \targ & \ctrl{-1} & \ctrl{1} & \targ & \qw & \ctrl{-1} & \targ & \qw & \targ & \ctrl{-1} & \qw & \qw\\
        \nghost{{q}_{6} :  } & \lstick{{q}_{6} :  } & \qw & \ctrl{-1} & \targ & \ctrl{-1} & \qw & \targ & \qw & \qw & \qw & \ctrl{-1} & \targ & \ctrl{-1} & \qw & \qw & \qw\\
        \nghost{{q}_{7} :  } & \lstick{{q}_{7} :  } & \gate{\mathrm{H}} & \qw & \ctrl{-1} & \qw & \qw & \qw & \qw & \qw & \qw & \qw & \ctrl{-1} & \qw & \qw & \qw & \qw\\
        \nghost{\mathrm{{c} :  }} & \lstick{\mathrm{{c} :  }} & \lstick{/_{_{8}}} \cw & \cw & \cw & \cw & \cw & \cw & \cw & \cw & \cw & \cw & \cw & \cw & \cw & \cw & \cw\\
    }}
    \caption{Quantum circuit for quantum random walk generation.}
    \label{fig:my-circuit}
\end{figure}

By inspecting the circuit, it can be observed that the structure is repeated twice. Qubits 0-3 handle the first 1D quantum walk, while qubits 4-7 handle the second one. Qubits 3 and 7 correspond to the coins, as the coins have been chosen to be the simplest operator possible to encode the act of tossing, which is Hadamard gates. Qubits 0-2 encode the $x$-axis, and 4-6 the $y$-axis. This circuit encodes a single step of a quantum random walk. For multiple steps, the same circuit must be concatenated as many times as the number of steps taken. The walk ends when measurements are performed. This circuit can be run on Qiskit backends. It is remotely executed and measured on IBM's simulator backend, and the results can be stored in a list for post-processing to construct the quantum paths in the classical rhythmspace.

\subsection{Mapping onto the rhythmspace interface}
\begin{figure}[H]
    \centering
    \includegraphics[scale=1]{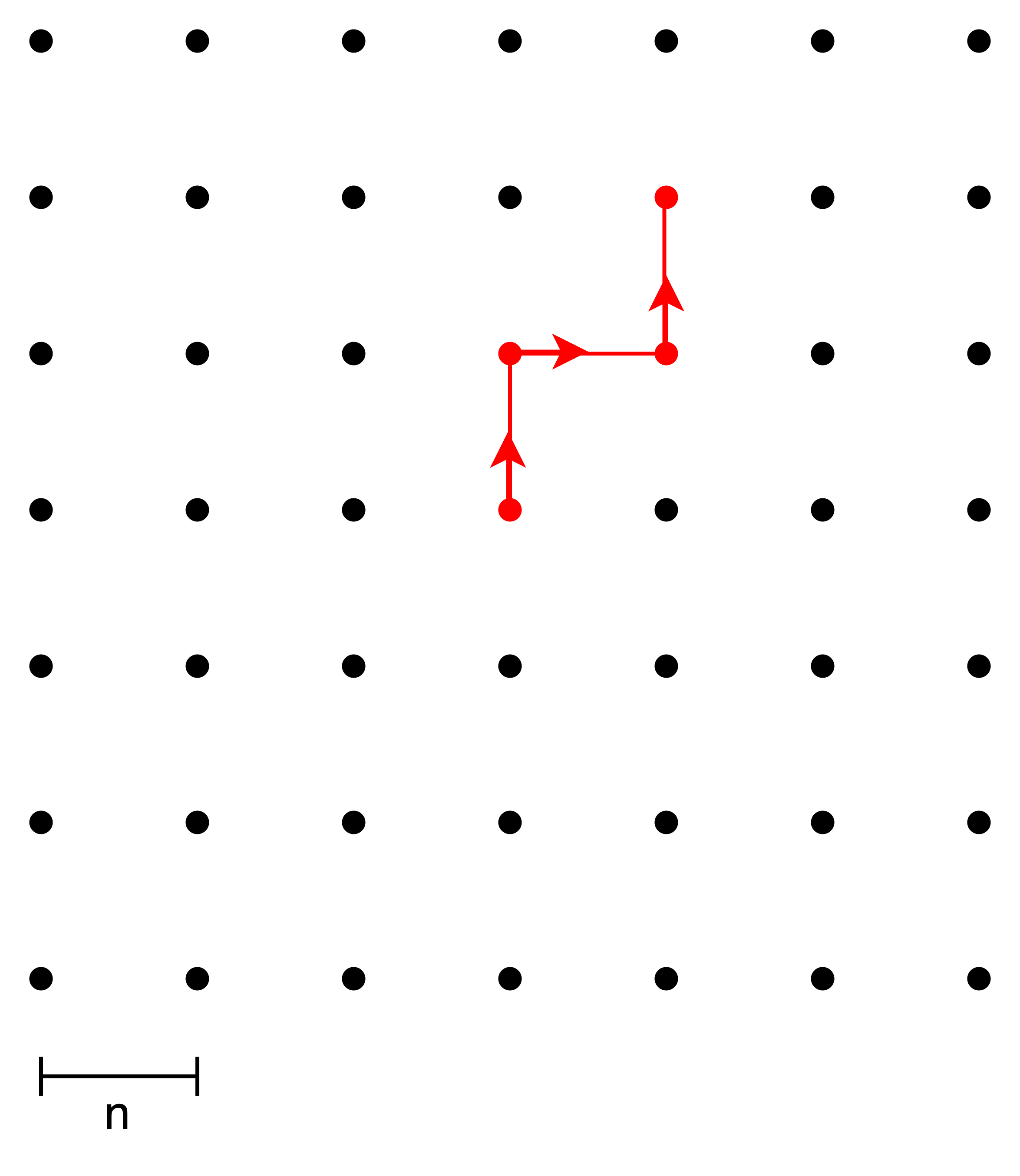}
    \caption{Illustration of the 49 positions of the quantum walk within the qubit space. An example of a 3-step walk is shown. The variable $n$ represents the pixels within a square of this qubit grid, and in this work, $n$ is set to 10.}
    \label{fig:grid}
\end{figure}

Once the encoding of positions as elements in a Hilbert space and the circuit are established, the mapping onto the rhythmspace must be designed. For this purpose, certain resolution aspects related to the rhythmspace interface must be addressed. Taking steps in a 2D quantum random walk implies that at every step, the space grid of size \( N \) grows as shown in Eq. (\ref{eq:grid-dimension}):

\begin{equation}
    N = (2T + 1) \cdot (2T + 1),
    \label{eq:grid-dimension}
\end{equation}

where \( T \) is the number of steps. Taking 3 steps creates a grid of 49 positions (see Fig. \ref{fig:grid}). Taking 4 steps would already surpass the available positions in the qubit space, since the walker space is formed by 6 qubits, and it has \( 2^6 = 64 \) positions. Thus, the quantum random walks are constrained to 3 steps, such that the qubit space has enough positions to encode it. The circuit must be repeated 3 times before measurement for it to encode 3 steps. After these 3 steps, the circuit is measured and sonified, which means that a rhythm is heard every 3 steps.

\begin{figure}[H]
    \centering
    \includegraphics[scale=1]{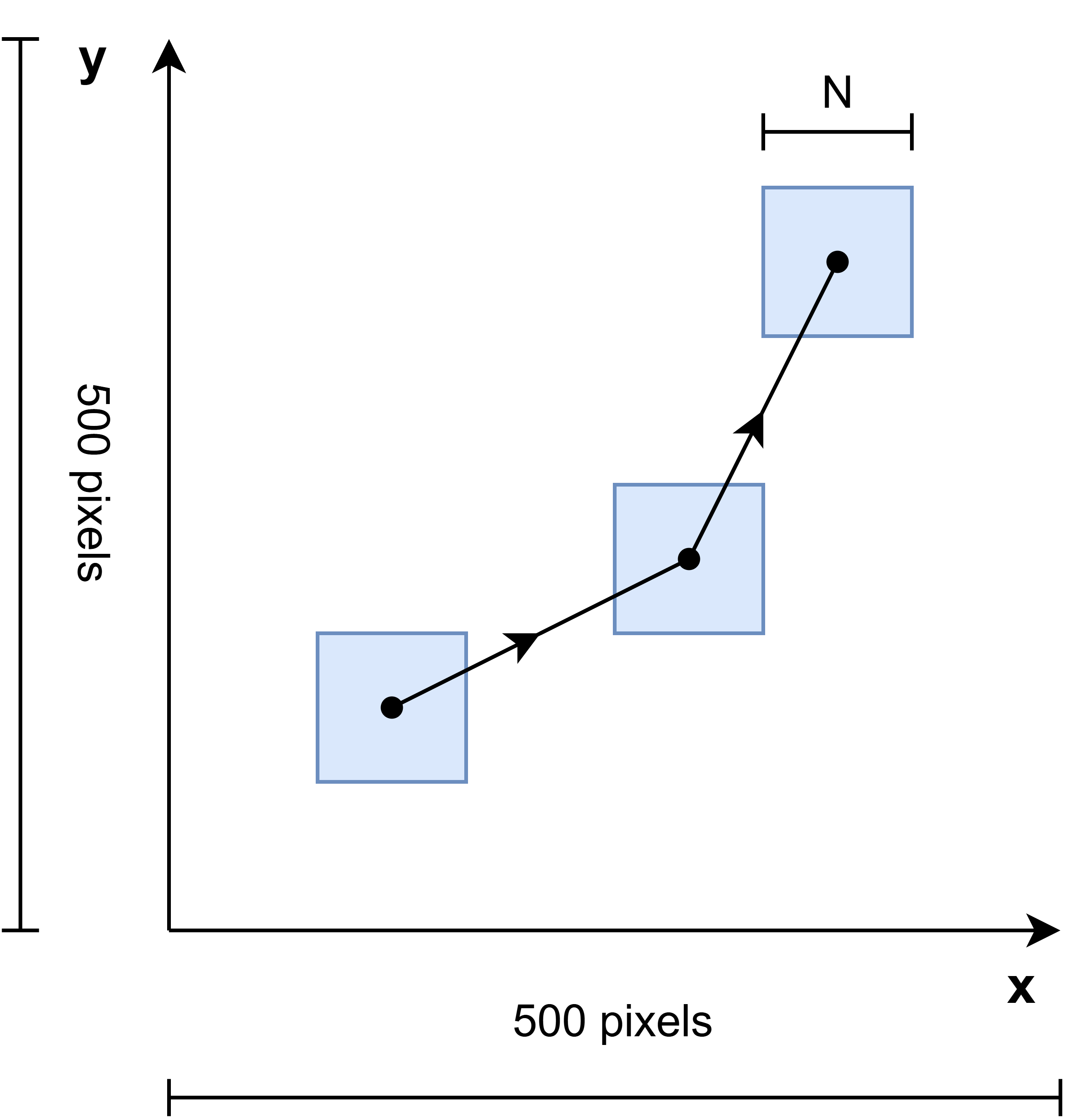}
    \caption{Qubits' window updating as the path in the rhythmspace progresses.}
    \label{fig:shift_window}
\end{figure}

On the other hand, rhythmspace is a 2D space system normalized between 0 and 1 in each dimension. In order to make it accessible for the user as a GUI, this 2D space must be formed by a square grid of pixels. In this paper, the grid is made up of 500 by 500 pixels. A correspondence between the pixel GUI in the rhythmspace and the quantum space must be defined. The 49 positions do not cover the entire grid of pixels in the rhythmspace. A solution to implement the quantum walks in this space consists of shifting and updating the window or grid of size \( N \) as the path progresses through the rhythmspace, such that it is always centered at the given position in this classical space (see Fig. \ref{fig:shift_window}). In addition, the pixels do not necessarily need to be mapped one-to-one to the positions in the walker Hilbert space. For example, there can be 3 pixels for every qubit square, which means that in the \( N =  7 \times 7 = 49 \) grid, there are \( (3 \times 6) \cdot (3 \times 6) \) pixels. A shifting window with a pixel space to Hilbert space ratio of $n=10$ was selected. This ratio was determined by computing the distances between all pairs of points in the rhythmspace used for interpolation. A significant number of these distances fall within the range of 0 to 0.2 (0-100 on the 500-pixel scale), indicating that a ratio between 0 and 100 pixels is reasonable.

Having established the implementation of 2D quantum walks on the rhythmspace interface, the analysis shifts to how classical potential fields can bias these paths, as explored in Section~\ref{sect:stage2}, enabling the generation of controlled or directed rhythmic evolutions.

\section{Second stage: biasing the quantum random walks through classical potential fields on the rhythmspace } \label{sect:stage2}
After the previous mappings are defined, the second stage generates the paths and plots them onto the interface. Potential fields dependent on the position $(x,y)$ of the rhythmspace are used here to bias the evolution of the quantum random walks.

\begin{figure}[H]
    \centering
    \includegraphics[scale=0.9]{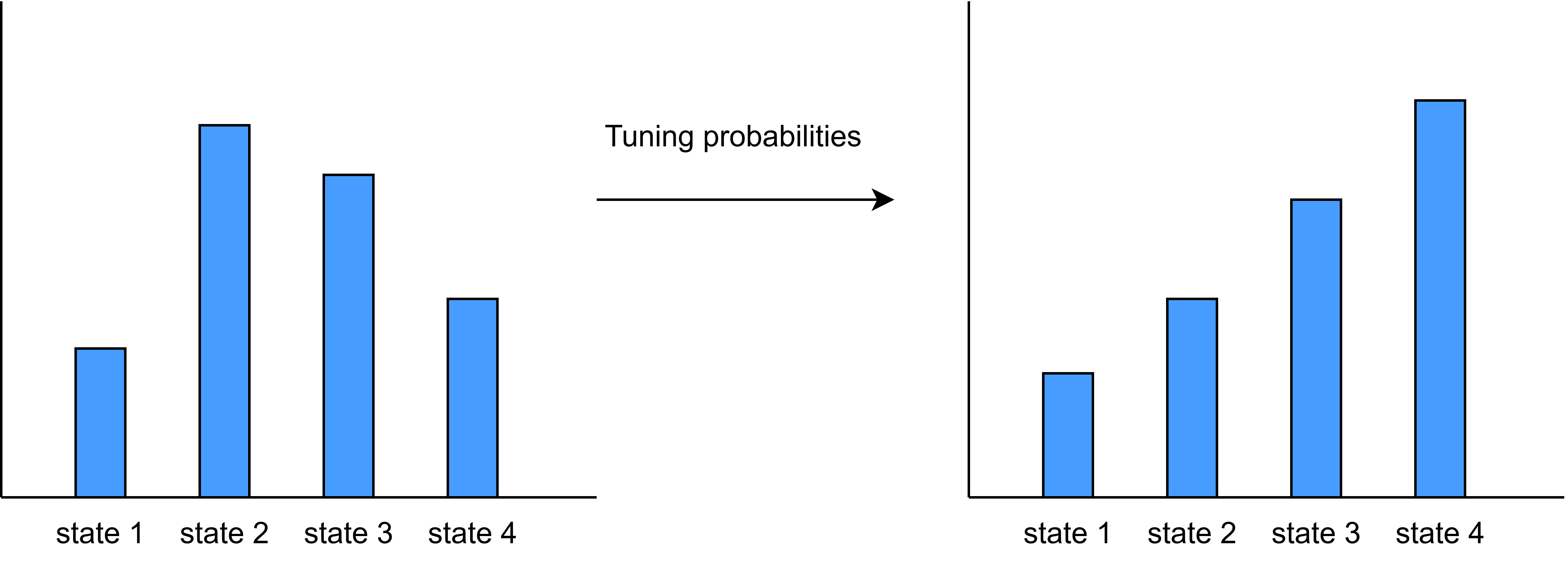}
    \caption{A simple diagram illustrating how the quantum feedback induced by a potential field tunes the probability distribution of the states of a given wave function. In this example, the tuning favors states 3 and 4, indicating that the potential is enhancing their corresponding positions in the rhythmspace.}
    \label{fig:tuning_probs}
\end{figure}

A fundamental characteristic of these potentials is their role in modifying the probability distribution of a quantum superposition state. In classical systems, the evolution of a trajectory is typically influenced by the gradient of the potential, which dictates the direction of motion through its derivative. However, in the quantum domain, how can a potential govern the evolution of a superposition of possible states? The solution lies in adjusting the probability distribution of wave functions by modifying the prepared states and circuit configurations. If the probability distribution defining the likelihood of different superposed directions is influenced by the gradient of the potential, then the potential effectively biases the quantum random walk. In other words, this mechanism induces a form of "quantum feedback," wherein the evolution of the quantum state under a potential field directly tunes the probability distributions of wave functions in the Hilbert space. A diagram illustrating this idea is shown in Fig. \ref{fig:tuning_probs}.

\begin{figure}[H]
    \centering
    \includegraphics[scale=0.84]{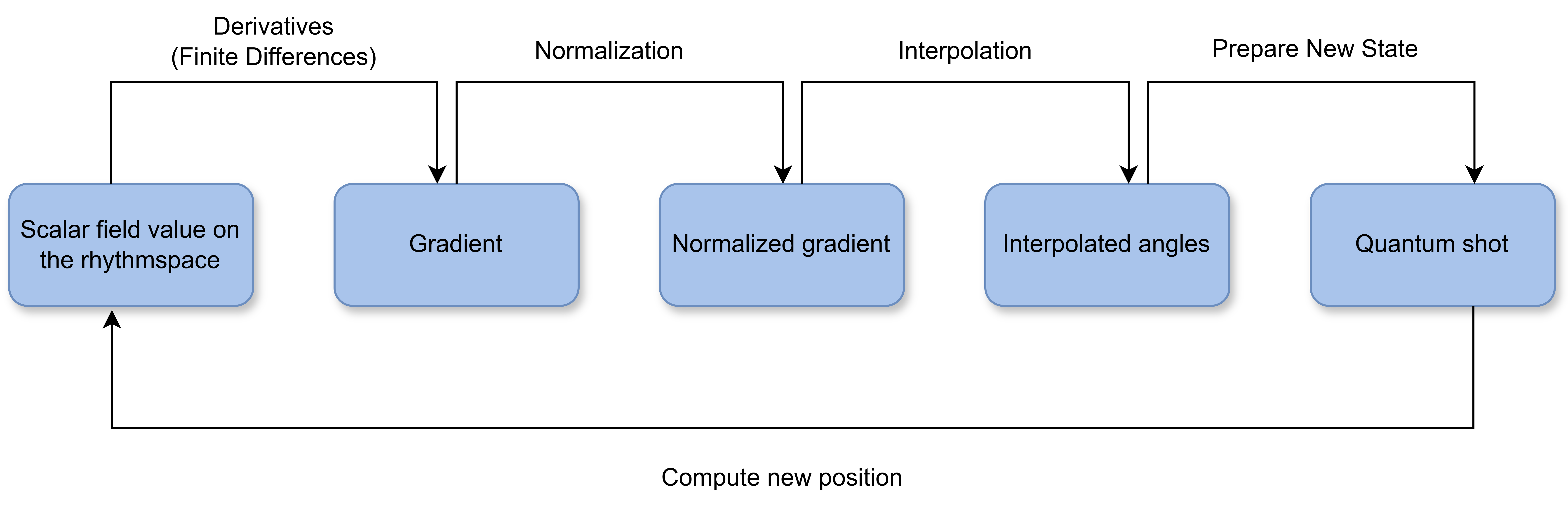}
    \caption{Quantum feedback algorithm steps.}
    \label{fig:generation_algorithm}
\end{figure}

In Fig. \ref{fig:generation_algorithm}, the steps followed by the algorithm to generate a new position are illustrated. The iterative algorithm takes the value of the potential field
$V$ at a point 
$(x,y)$ within the rhythmspace grid. It then computes the gradient vector of this potential by applying the finite differences method with respect to the two variables. The gradient tuple values are normalized to the range 0-1.

Once normalized, the gradient tuple is used to interpolate the values of the angles required to prepare the new state vector to be input into the quantum circuit. This interpolation follows the expressions given in Eq. \eqref{eq:angles}: 

\begin{align}
    \Phi_{i,x} = (1-t_x) \Phi_{min}+ t_x\Phi_{max}\notag\\
    \Theta_{i,x} = (1-t_x) \Theta_{min}+ t_x\Theta_{max}\notag\\
    \Phi_{i,y} = (1-t_y) \Phi_{min}+ t_y\Phi_{max}\notag\\
    \Theta_{i,y} = (1-t_y) \Theta_{min}+ t_y\Theta_{max},
    \label{eq:angles}
\end{align}

where $(t_x,t_y)$ is the normalized gradient tuple. These angles are used to apply rotations to the coin qubit in each 1D quantum random walk within the main circuit. There are two angles, $\Phi$ and $\Theta$, per axis. These rotations alter the initial state vector in such a way that the probability distribution of the possible outputs, after measuring the wave function, is modified. Essentially, the potential field tunes the possible outputs of the circuit at each step. Although the same four possible outputs can be measured, the probability of each one changes at every step. The potential field directly biases the probabilities, creating a feedback loop between the gradient and the probability distributions in the circuit.

To interpolate the angle values in Eq. (\ref{eq:angles}), a classical optimization through gradient descent is performed on the circuit, defining a cost function to establish the boundaries of the angles $\Phi_{min}, \Phi_{max}, \Theta_{min}$ and $\Theta_{max}$.

Different potential field scenarios that include the quantum feedback algorithm are explored. The following subsections introduce them one by one.

\subsection{Null potential field}
\begin{figure}[H]
    \centering
    \includegraphics[scale=0.6]{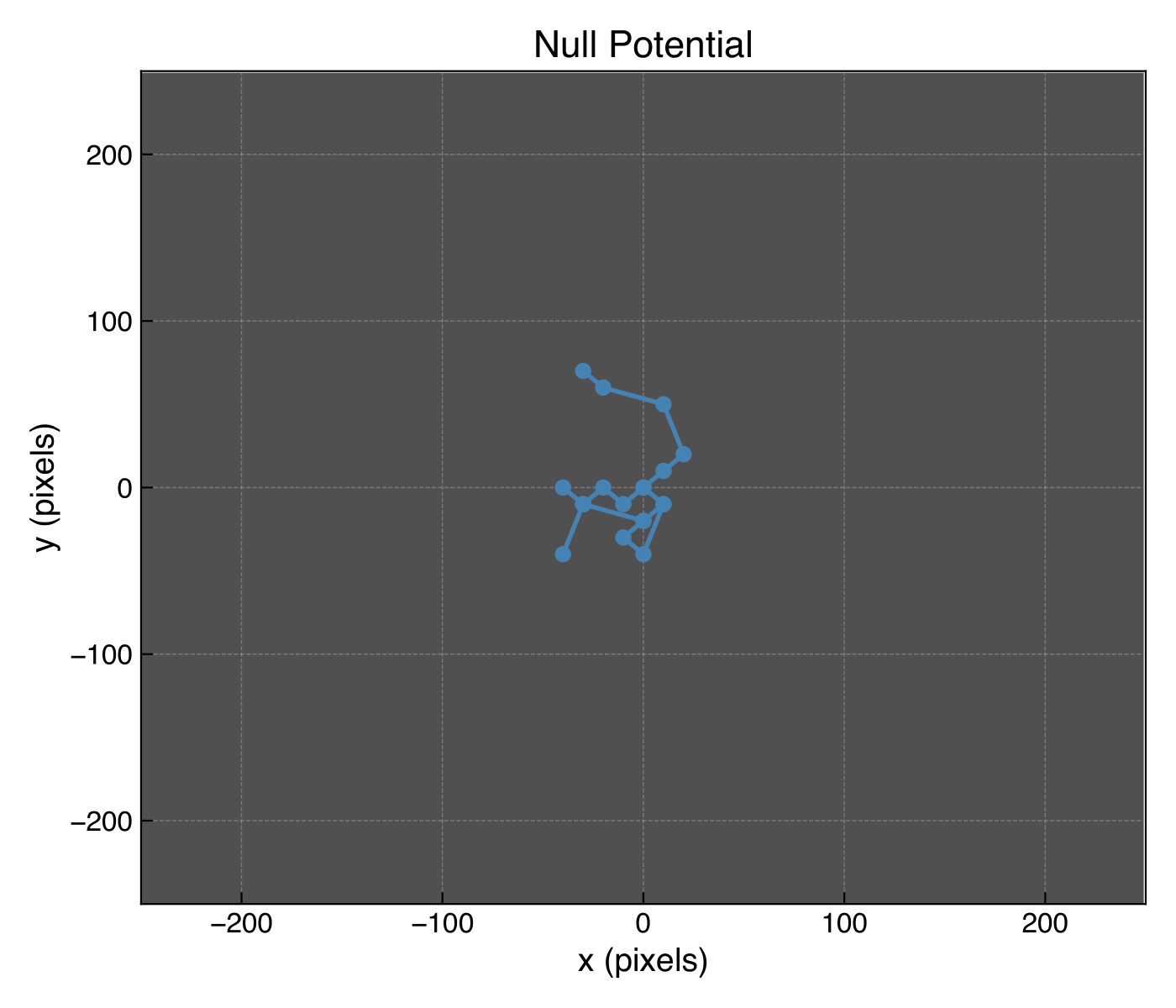}
    \caption{Quantum random walk path. Since the potential is zero, no bias is induced in any specific direction, and the walk just wanders around without a specific pattern. The initial position is $(0,0)$.}
    \label{fig:null_potential}
\end{figure}

In the first case, the potential is zero, and there is no position-dependent gradient to bias the probability distribution of the quantum walks in a given direction, as shown in Fig. \ref{fig:null_potential}. This is the simplest scenario where all directions are equally probable.

\subsection{Linear potential field}
In the second case, a linear potential induces paths that follow a line towards the minimum energy regions in the space. In this case, the quantum feedback biases the probability distribution towards a given direction. Notice that a classical path would strictly follow a line. However, a quantum path introduces a certain probability at every given step of following different directions. The key idea is that the probability of moving in the direction imposed by the potential is higher, but the probabilistic nature still introduces chaos and randomness, making the sonification special and interesting. Eq. \eqref{eq:linear-potential} shows the expression for the linear potential:

\begin{equation}
    V(y) = ky.
    \label{eq:linear-potential}
\end{equation}

\begin{figure}[H]
    \centering
    \includegraphics[scale=0.6]{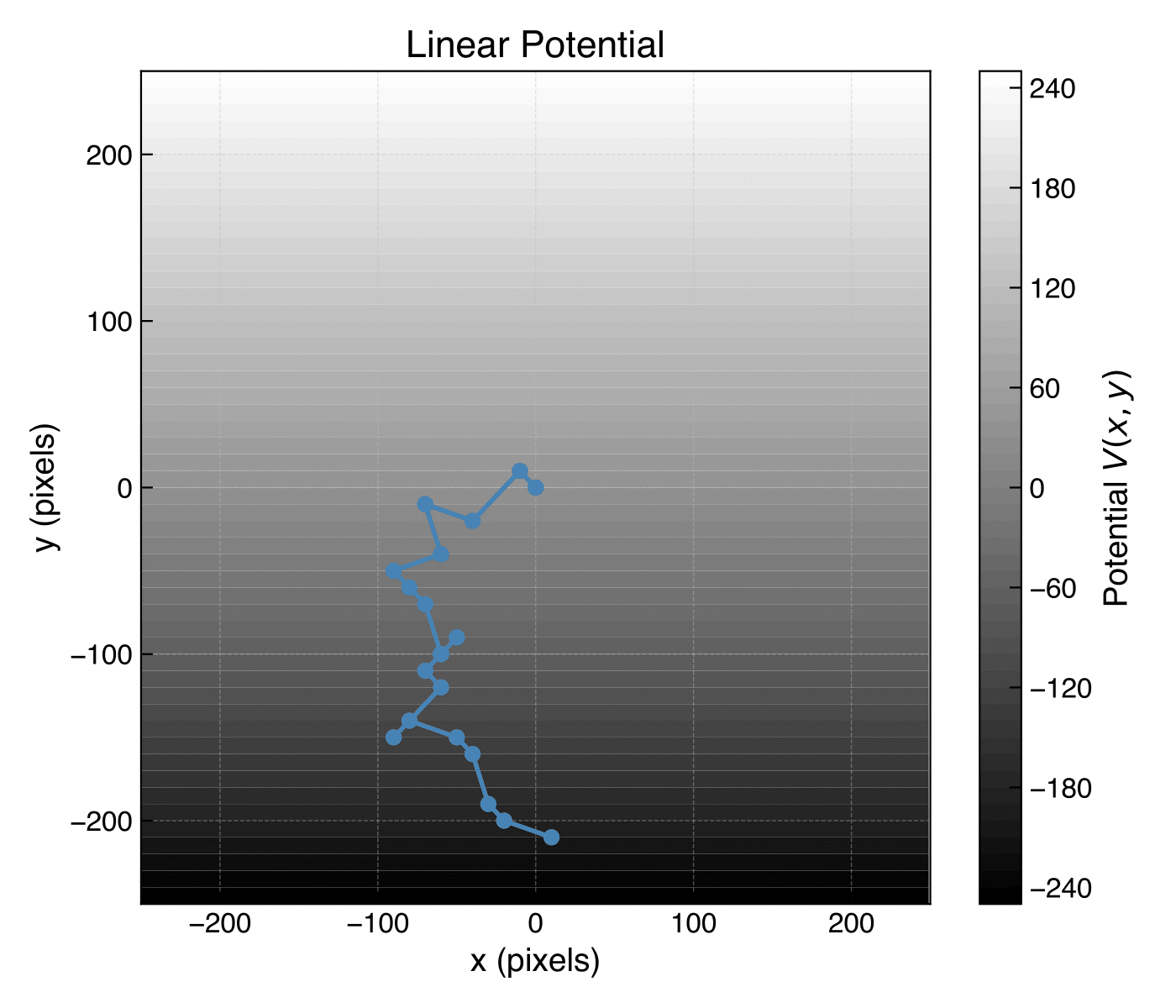}
    \caption{Path biased along the y-axis. Since it is a quantum random walk, small deviations at every step can be observed as a result of the probability distributions of wave functions. However, the potential bias still dominates, inducing directionality. The initial position is $(0,0)$ and the slope is $k=1$.}
    \label{fig:linear_potential}
\end{figure}

In this work, $k$ is set to 1. Fig. \ref{fig:linear_potential} illustrates an example of a path under this potential.

\subsection{Gaussian potential field}
\begin{figure}[H]
    \centering
    \includegraphics[scale=0.6]{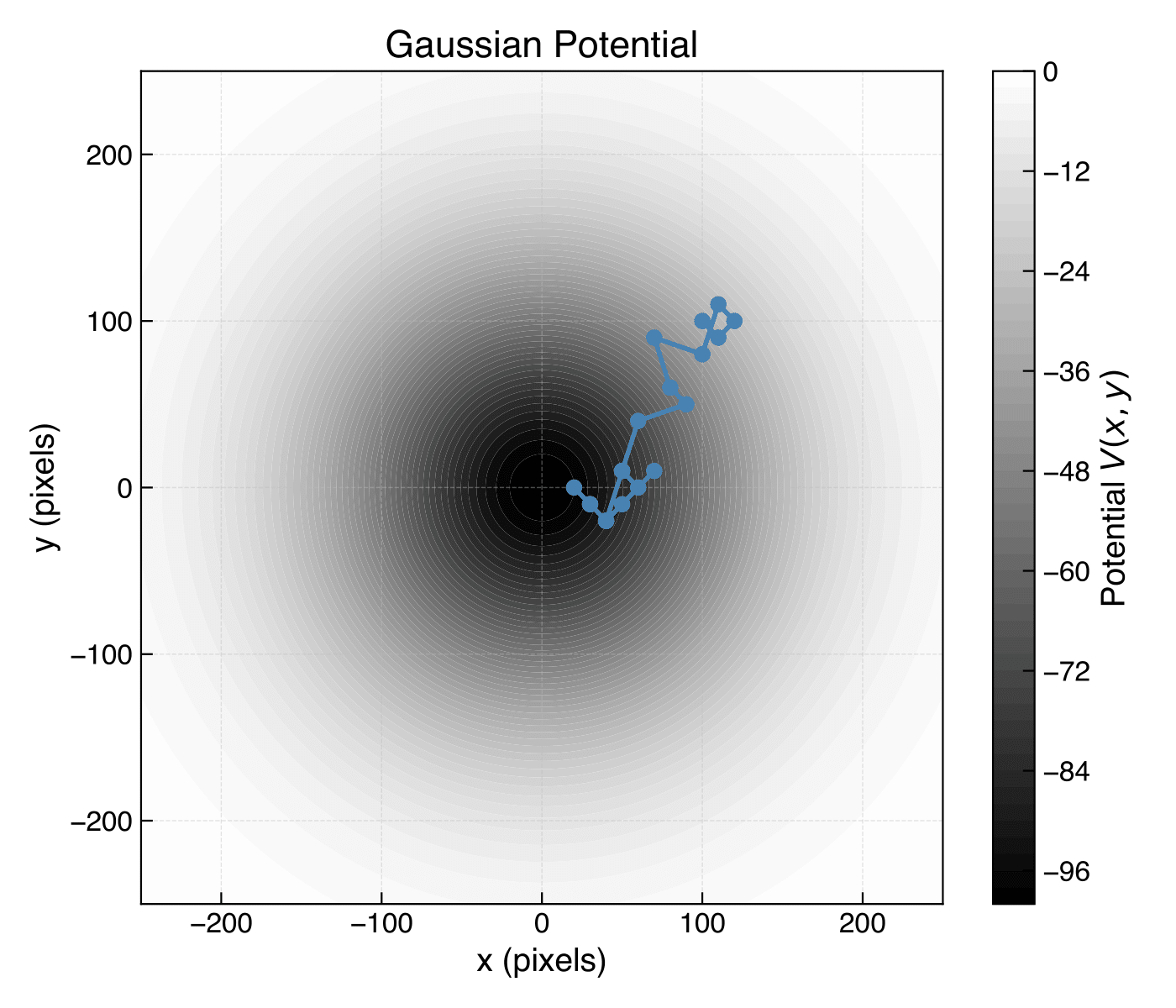}
    \caption{The Gaussian potential creates a well into which the path falls when approached. Despite being a quantum walk, the probability distribution is unable to escape the well due to the higher weight privileging the gradient descent. The initial position is $(100,100)$, and the variables of the potential field are $V_0 = 100$, \(x_0 = y_0 = 0\) and \(\sigma_x = \sigma_y = 100\).}
    \label{fig:gaussian_potential}
\end{figure}

The third case is a Gaussian potential that sets a potential well. If the path approaches the well, it will progressively fall into it. The fall will not be direct, since there is a certain probability of avoiding it at each step, but the bias will always favor advancement towards the deepest region of the well. The Gaussian potential in a 2D \(x-y\) plane can be expressed as shown in Eq. \eqref{eq:gaussian-potential}:

\begin{equation}
    V(x, y) = V_0 \exp\left(-\frac{(x - x_0)^2}{2\sigma_x^2} - \frac{(y - y_0)^2}{2\sigma_y^2}\right),
    \label{eq:gaussian-potential}
\end{equation}

where, \(V_0\) is the amplitude of the potential, \(x_0, y_0\) are the coordinates of the center of the Gaussian potential, and \(\sigma_x, \sigma_y\) are the standard deviations in the \(x\) and \(y\) directions, respectively. In this work, \(V_0\) has been chosen to be 100, \(x_0 = y_0 = 0\), and \(\sigma_x = \sigma_y = 100\). Fig. \ref{fig:gaussian_potential} shows an example of a path generated under this potential.

\subsection{Gaussian potential field under inertial dynamics}
So far, quantum feedback does not account for accelerations. If inertia is included within the equations, second-order paths are induced. This is the case of the last potential used. The equations used for the non-inertial case update the positions as follows in Eq. \eqref{eq:positions-nonintertial-potential}:

\begin{align}
    x' &= x + \Delta x \notag\\
    y' &= y + \Delta y,
    \label{eq:positions-nonintertial-potential}
\end{align}

where the accelerations are zero and the positions evolve through a delta. The inertial case updates positions as it is given in Eq. \eqref{eq:positionsvelocities-intertial-potential}:

\begin{align}
    v_x' &= v_x + a_x \Delta t, \;
    v_y' = v_y + a_y \Delta t \notag\\
    x' &= x + v_x' \cdot \Delta t, \;
    y' = y + v_y' \cdot \Delta t.
    \label{eq:positionsvelocities-intertial-potential}
\end{align}

\begin{figure}[H]
    \centering
    \includegraphics[scale=0.6]{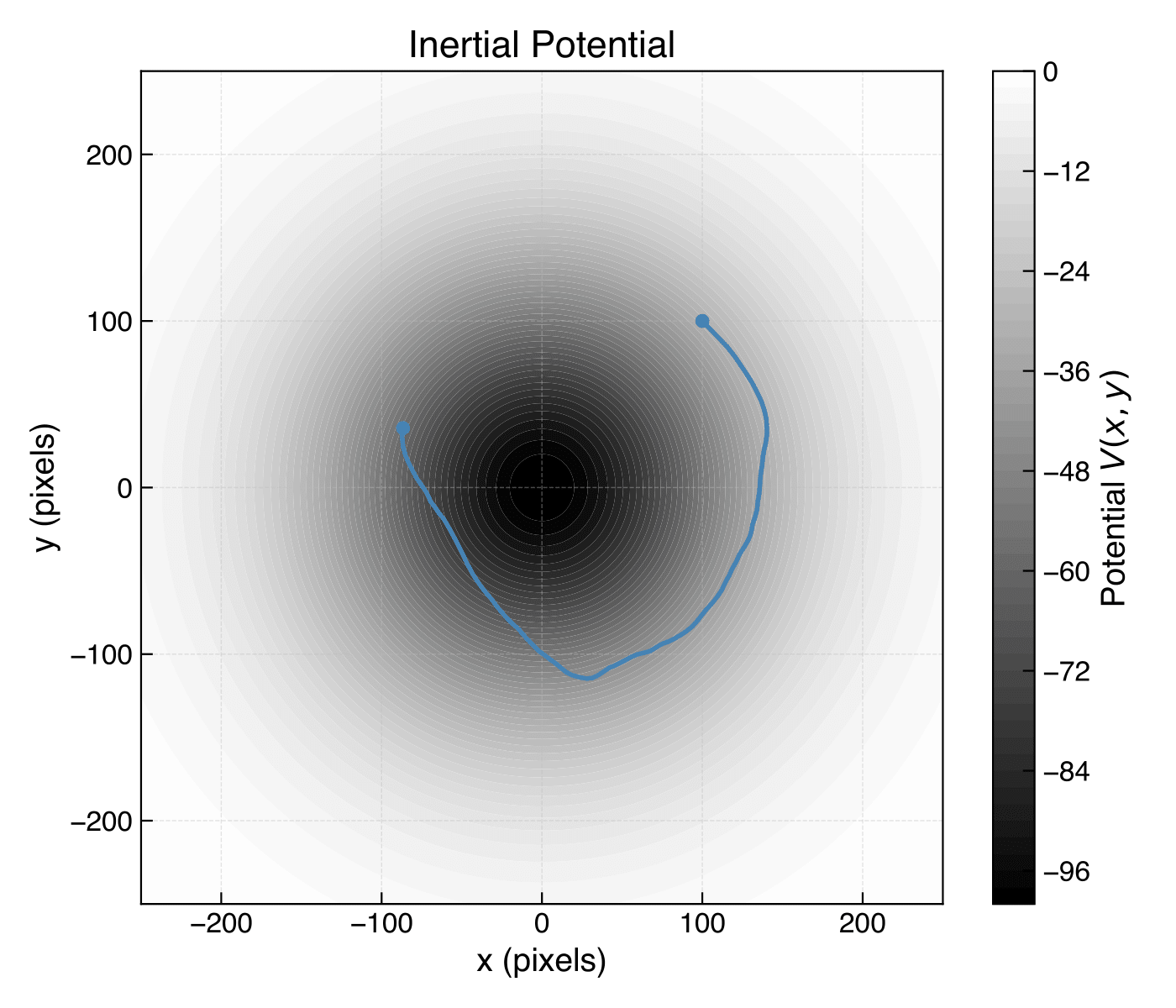}
    \caption{As the path approaches the well, it gains inertia and starts drawing a curve around it. In classical conditions, the path would describe a perfectly symmetrical orbit. Nevertheless, the quantum walk perturbs the orbit, introducing quantum randomness that deforms it. The initial position is $(100,100)$, the initial velocities are $(10,-10)$, and the acceleration variables are $m=10$ and $\Delta t = 0.52$.}
    \label{fig:inertial_potential}
\end{figure}

Here, the potential bias controls the accelerations instead of doing it directly on the positions. Notice that quantum paths do not draw perfect closed orbits around the Gaussian well; instead, they have a certain probability of altering the orbit at each step. If the bias is subtle enough, the quantum path will be able to draw a pseudo-closed orbit, which resembles a classical one but introduces variations. Fig. \ref{fig:inertial_potential} shows an example of this case. 

Now that the plotting onto the interface introducing the potential fields has been handled, the following section, Section \ref{sect:stage3}, explains how the paths are sonified using the drum pattern generation of the rhythmspaces.

\section{Third stage: sonification} \label{sect:stage3}
The final stage of the methodology involves sonifying the quantum walks by encoding them as MIDI messages, which are then converted to audio using the generative algorithm of the rhythmspace. The resulting audio files can be exported from the DAW and added to animated videos of the trajectories.

As each point in the rhythmspace outputs a rhythm pattern, a quantum path describes a series of 2D positions that each can become a drum pattern. In order to provide a coherent rhythmic progression, the path is resampled onto a 16th semiquaver grid in this case, ensuring that the rhythmic evolution is structured. Once the resampled path is computed, the rhythmic patterns corresponding to each point are generated. To build the final rhythmic sequence, a single slice from each pattern at each resampled dot is selected.

\begin{figure}[H]
    \centering
    \includegraphics[scale=1.5]{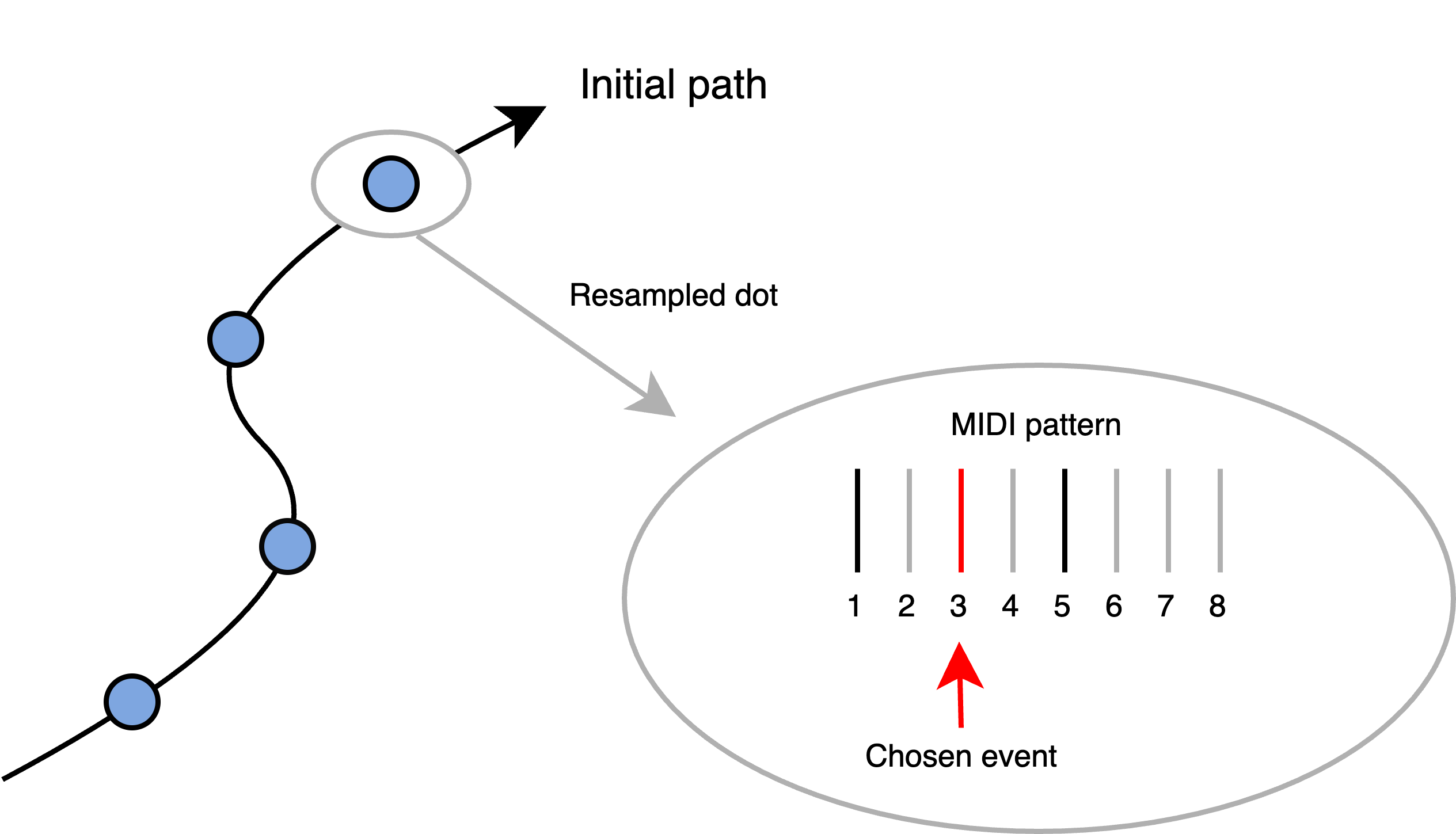}
    \caption{The black line represents the initial quantum random walk path. The blue dots indicate the resampled path that aligns with the imposed semiquaver grid for the rhythm. Each resampled dot contains a MIDI pattern. The event corresponding to the specific moment in the grid is extracted from the pattern and used to construct the global drum pattern. This figure illustrates an 8-semiquaver grid.}
    \label{fig:sonification}
\end{figure}

The elements or slices within a pattern are called "events". The event chosen within each pattern is the one that aligns with the global beat of the grid. An event within a pattern consists of a list of percussive instruments to be triggered at that instant. Therefore, all the selected events contain the information about which instruments should be triggered at each instant of the temporal percussive grid. Fig. \ref{fig:sonification} illustrates this resampling process.

The lists of instruments to be triggered, corresponding to the selected events, are encoded in the MIDI protocol and sent to a DAW to produce sound. The MIDI messages received in the DAW create percussive patterns that trigger different instruments at each beat. Therefore, the sequence of patterns must align with the DAW's temporal grid. The MIDI messages must be sent at time intervals that correspond to this grid.

Once the MIDI messages arrive at the DAW, they are sonified using a sampler that follows the General MIDI Percussion Map labeling \cite{generalmidi2024}. By convention, different MIDI notes correspond to different percussion instruments. Thus, each MIDI string triggers a specific set of instruments at a given time. The following list wraps up this process step by step:

\begin{enumerate}
    \item Gather measured path data from the quantum random walk.
    \item Resample the path to match a desired temporal grid.
    \item Map each point in the rhythmspace to a stored drum pattern.
    \item Choose the correct event (i.e., the current element or slice) of that pattern.
    \item Send MIDI messages to the DAW in sync with the chosen global timeline.
\end{enumerate}

Examples of various sonified quantum random walks in a rhythmspace under different potential field conditions can be visualized and heard here: \href{https://www.youtube.com/playlist?list=PL50U6FDg9kH3QhYTJYPfgDqk2LIZypS2m}{\textit{Click here to watch the videos}} \cite{aguadoyanez2025}. These animations were generated by exporting the audios from Cubase and attaching them to the videos created with Python.

By sonifying the paths and matching the generated audio with the animated videos, the entire process of generating quantum random walk paths sonified and visualized using the rhythmspaces interface is completed. Section \ref{sect:disc} discusses the results achieved across the three stages presented in the previous sections.

\section{Discussion} \label{sect:disc}

This work details the development of more complex proof-of-concept quantum random walk algorithms for sonification purposes. It presents a method that exploits the symmetries of the problem to construct robust and efficient algorithms that minimize the number of qubits and gates used, taking into account the current limitations of existing hardware. These algorithms are also scalable, allowing the procedures to be easily extended to larger qubit systems. For example, the mappings between the Hilbert space and the rhythmspace can naturally accommodate higher-dimensional Hilbert spaces, and the windows and resolutions can be rescaled to suit specific problems. The algorithms presented successfully expand the techniques in the references to a more complex geometry, in this case, a 2D plane. Since quantum computing is still limited by the current existing and available hardware, for future work, it would be interesting to expand these algorithms when more powerful computers become available and to rethink them in this new scenario.

The field of quantum computing has properties that are inherently different in nature from those of classical computers. Hence, the music generated with quantum computers relies on a different mathematical formalism. Unlike classical walks, the quantum random walks generated in this work exploit superposition and entanglement, which can lead to distinctive rhythmic variation. When more powerful computers become available, the algorithms developed could be extended to further explore these properties. For example, measurements could be performed after more steps, potentially even hundreds, instead of just three, to amplify the quantum computing effects. Additionally, more qubits and gates could be employed to enhance these properties.

Regarding the GUI, currently interaction requires accessing the code and manually changing the parameters for the quantum paths and potentials. Ideally, this project would evolve to allow users to explore these parameters through accessible buttons and sliders on an interface. On the other hand, the visualization of the effects of these parameters on the paths is already being plotted and displayed through animations. The main area needing improvement is the method for modifying parameters in the code, which would benefit from a more immediate and interactive approach, but the resolution and plotting aspects have already been addressed.

It is worth reiterating the significance of the concept of quantum feedback, used to bias the paths through potential fields. This concept directly connects the visual interface and audio with the algorithms in a creative way and could be exploited in real-time performance scenarios. The idea of manipulating an interface that somehow affects the wave functions of the qubits is a promising approach for exploring quantum algorithms through art.

Researchers can use approaches like this one to better understand abstract quantum computing algorithms, particularly quantum random walks, by encoding information with qubits for artistic purposes, which bridges art and science. Smartly designed sonification mappings can help trigger the right intuitions through sound, making abstract concepts more accessible. To evaluate whether these algorithms effectively deliver such ideas, user-based studies could be conducted, adjusting potential fields and algorithms to make quantum perturbations both audible and visually meaningful. These studies could reveal how both researchers and the general public interpret and engage with complex physics through music.

As a final appreciation, notice that a rhythmspace is a multidimensional space. Various other sound datasets could be arranged and interpolated to create different types of spaces. The quantum algorithms in this work can operate on any sound space; they are not strictly bound to a rhythmspace. It can also be viewed as a method for encoding and decoding audio, similar to the function of a variational autoencoder \cite{variationalautoencoder2025}. In this context, the rhythmspace serves as the latent space. This analogy suggests the potential for connecting quantum random walks in audio with Machine Learning algorithms.

\section{Conclusion}
This work introduces a novel interactive, scalable, and efficient quantum computing algorithm for rhythm generation, extending the use of quantum random walks in music to 2D rhythmspace interfaces that users can explore. By decomposing 2D walks into two 1D components, the circuit depth was minimized. Through the incorporation of classical potential fields, various shapes of quantum trajectories can be generated, allowing for creative interaction by biasing the evolving direction of the paths. These paths are then sonified into MIDI drum patterns, which are rendered into audio within a DAW. This work serves both as a proof of concept for quantum computing-based music generation and as a foundation for interdisciplinary exploration between quantum physics and the arts. Since these quantum-based methods differ from purely classical random generation, new possibilities and insights for artistic music generation arise. As quantum hardware improves, the algorithms could be expanded to explore more complex quantum behavior, enabling longer walks, more qubits, and richer audio and visual representations. 

\acknowledgement{The authors would like to acknowledge the support from the Music Technology Group of Universitat Pompeu Fabra, and from the Center for Quantum Technology and Applications of DESY supported by the funds from the Ministry of Science, Research and Culture of the State of Brandenburg.

\begin{center}
\centerline{
    \includegraphics[width=0.1\textwidth]{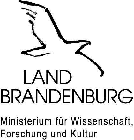}}
\end{center}}


    \authorcontributions{software: María Aguado-Yáñez and Daniel Gómez-Marín; writing: María Aguado-Yáñez; methodology, reviewing and editing: all authors.}

\availabilityofdataandmaterials{The data that support the findings of this study are openly available in the rhythmspaces repository at \cite{gomezmarin2024}.}

\reftitle{References}
\bibliography{main}

\begin{thebibliography}{10}

\bibitem{quantumcomputing2025}
{IBM}. {IBM}, editor. What Is Quantum Computing?. IBM; 2025.
\newblock Accessed: 3 March 2025.
\newblock Available from: \url{https://www.ibm.com/think/topics/quantum-computing}.

\bibitem{miranda2022}
Miranda ER.
\newblock Quantum Computer Music: Foundations, Methods and Advanced Concepts.
\newblock Cham: Springer; 2022.

\bibitem{miranda2024}
Miranda ER, editor.
\newblock Advances in Quantum Computer Music.
\newblock World Scientific; 2024.

\bibitem{venegas2012}
Venegas-Andraca SE.
\newblock Quantum Walks: A Comprehensive Review.
\newblock Quantum Information Processing. 2012;11(5):1015-106.

\bibitem{feynman1965}
Feynman RP, Leighton RB, Sands M.
\newblock The Feynman Lectures on Physics.
\newblock Reading, MA: Addison-Wesley; 1965.

\bibitem{turquois2016}
Turquois C, Hermant M, Gómez-Marín D, Jordà S.
\newblock Exploring the benefits of 2D visualizations for drum samples retrieval.
\newblock In: Proceedings of the 2016 ACM Conference on Human Information Interaction and Retrieval. ACM; 2016. p. 329-32.

\bibitem{gomezmarin2018}
G\'omez-Mar\'in D. G\'omez-Mar\'in D, editor. Similarity and Style in Electronic Dance Music Drum Rhythms. Barcelona, Spain: Universitat Pompeu Fabra; 2018.
\newblock PhD Thesis, Universitat Pompeu Fabra.
\newblock PhD Thesis.

\bibitem{gomezmarin2020}
Gómez-Marín D, Jordà S, Herrera P.
\newblock Drum rhythm spaces: From polyphonic similarity to generative maps.
\newblock Journal of New Music Research. 2020;49(5):438-56.

\bibitem{gomezmarindemo2018}
G\'omez-Mar\'in D. G\'omez-Mar\'in D, editor. RhythmSpace: EDM Demo. YouTube; 2018.
\newblock Accessed: 3 March 2025.
\newblock YouTube video.
\newblock Available from: \url{https://www.youtube.com/watch?v=Wwg2XqK4vKQ}.

\bibitem{mirandabasak2022}
Miranda ER, Basak S.
\newblock Quantum computer music: Foundations and initial experiments.
\newblock In: Miranda ER, editor. Quantum Computer Music: Foundations, Methods and Advanced Concepts. Cham: Springer International Publishing; 2022. p. 43-67.

\bibitem{allen2022}
Allen E, Bulmer J, Small S.
\newblock Making music using two quantum algorithms.
\newblock In: Miranda ER, editor. Quantum Computer Music: Foundations, Methods and Advanced Concepts. Cham: Springer International Publishing; 2022. p. 69-82.

\bibitem{yamada2023}
Yamada R, Piñol E, Grandi S, Zakrzewski J, Lewenstein M.
\newblock Towards the Intuitive Understanding of Quantum World: Sonification of Rabi Oscillations, Wigner functions, and Quantum Simulators.
\newblock arXiv preprint arXiv:231113313. 2023.
\newblock [cited 31 March 2025].

\bibitem{crippa2024}
Crippa A, Chai Y, Hamido OC, Itaborai P, Jansen K.
\newblock Quantum Computing Inspired Paintings: Reinterpreting Classical Masterpieces.
\newblock arXiv preprint arXiv:241109549. 2024.
\newblock [cited 31 March 2025].

\bibitem{dmytrofedoriaka2020}
Fedoriaka D. Fedoriaka D, editor. Decomposing Unitary Matrix Into Quantum Gates. GitHub; 2020.
\newblock Accessed: 3 March 2025.
\newblock GitHub repository.
\newblock Available from: \url{https://github.com/fedimser/quantum_decomp/blob/master/example.ipynb}.

\bibitem{venegas2023}
Wing-Bocanegra A, Venegas-Andraca SE.
\newblock Circuit Implementation of Discrete-Time Quantum Walks via the Shunt Decomposition Method.
\newblock Quantum Information Processing. 2023;22(3):146.

\bibitem{generalmidi2024}
{MIDI Association}. {MIDI Association}, editor. General MIDI Level 1. MIDI Association; 2024.
\newblock Accessed: 4 March 2025.
\newblock Available from: \url{https://midi.org/general-midi-level-1}.

\bibitem{aguadoyanez2025}
Aguado-Y\'a\~nez M. Aguado-Y\'a\~nez M, editor. Quantum Random Walks on Rhythmspaces. YouTube; 2025.
\newblock Accessed: 7 April 2025.
\newblock YouTube playlist.
\newblock Available from: \url{https://www.youtube.com/playlist?list=PL50U6FDg9kH1pdHPTLMwotoPtsr7ttYQu}.

\bibitem{variationalautoencoder2025}
{IBM}. {IBM}, editor. What Is a Variational Autoencoder?. IBM; 2024.
\newblock Accessed: 4 March 2025.
\newblock Available from: \url{https://www.ibm.com/think/topics/variational-autoencoder}.

\bibitem{gomezmarin2024}
G\'omez-Mar\'in D, Aguado-Y\'a\~nez M. G\'omez-Mar\'in D, Aguado-Y\'a\~nez M, editors. rhythmspaces: quantum-rhythms branch. GitHub; 2024.
\newblock Accessed: 31 March 2025.
\newblock GitHub repository.
\newblock Available from: \url{https://github.com/danielgomezmarin/rhythmspaces/tree/quantum-rhythms}.

\end{thebibliography}



\end{document}